\documentclass{emulateapj}
\usepackage{graphicx}

\def\refjnl#1{{\rm#1}}
\def\apjl{\refjnl{ApJ}}                
\def\apjs{\refjnl{ApJS}}               
\def\ao{\refjnl{Appl.~Opt.}}           
\def\aap{\refjnl{A\&A}}                

\def\be{\begin{equation}} 
\def\ee{\end{equation}}

\def\gsim{\lower.5ex\hbox{\gtsima}} 
\def\lsim{\lower.5ex\hbox{\ltsima}} \def\gtsima{$\; \buildrel > \over \sim \;$} \def\ltsima{$\; \buildrel < \over \sim \;$} \def\prosima{$\; 
\buildrel \propto \over \sim \;$} \def\gsim{\lower.5ex\hbox{\gtsima}} 
\def\lsim{\lower.5ex\hbox{\ltsima}} 
\def\simgt{\lower.5ex\hbox{\gtsima}} 
\def\simlt{\lower.5ex\hbox{\ltsima}} 
\def\simpr{\lower.5ex\hbox{\prosima}}   
  
 \def\gtsima{$\; \buildrel > \over \sim \;$} 
\def\ltsima{$\; \buildrel < \over \sim \;$} 
\def\gsim{\lower.5ex\hbox{\gtsima}} 
\def\lsim{\lower.5ex\hbox{\ltsima}} 
\def\simgt{\lower.5ex\hbox{\gtsima}} 
\def\simlt{\lower.5ex\hbox{\ltsima}} 
\def\simpr{\lower.5ex\hbox{\prosima}}

\def\E3{{\cal E}_{\rm g}^{III}}

\def\Msun{\rm M_\odot}
\def\Zsun{\rm Z_\odot}

\def\Msun{\rm M_\odot}
\def\Zsun{\rm Z_\odot}
\def\M*{M_*}
\def\Z*{Z_*}
\def\L*{L_*}
\def\MUV{M_{UV}}

\def\kev{\,\rm{keV}} 
\def\mx{\,m_x} 
\def\highz{high-$z$\,} 

\shorttitle{Early galaxies in WDM models}
\shortauthors{Dayal et al.}

\begin{document}

\title{Early galaxy formation in warm dark matter cosmologies}
\author{Pratika Dayal\altaffilmark{1}, Andrei Mesinger\altaffilmark{2} \& Fabio Pacucci\altaffilmark{2}}
\altaffiltext{1}{Institute for Computational Cosmology, Department of Physics, University of Durham, South Road, Durham DH1 3LE, UK}
\altaffiltext{2} {Scuola Normale Superiore, Piazza dei Cavalieri 7, 56126 Pisa, Italy }

\begin{abstract}
We present a framework for high-redshift ($z \gsim 7$) galaxy formation that traces their dark matter (DM) and baryonic assembly in four cosmologies: Cold Dark Matter (CDM) and Warm Dark Matter (WDM) with particle masses of $\mx =$ 1.5, 3 and 5 $\kev$. We use the same astrophysical parameters regulating star formation and feedback, chosen to match current observations of the evolving ultra violet luminosity function (UV LF). We find that the assembly of observable (with current and upcoming instruments) galaxies in CDM and $\mx \geq 3 \kev$ WDM results in similar halo mass to light ratios (M/L), stellar mass densities (SMDs) and UV LFs. However the suppression of small-scale structure leads to a notably delayed and subsequently more rapid stellar assembly in the $1.5 \kev$ WDM model. Thus galaxy assembly in $\mx \lsim 2\kev$ WDM cosmologies is characterized by: (i) a dearth of small-mass halos hosting faint galaxies; and (ii) a younger, more UV bright stellar population, for a given stellar mass. The higher M/L ratio (effect ii) partially compensates for the dearth of small-mass halos (effect i), making the resulting UV LFs closer to CDM than expected from simple estimates of halo abundances. We find that the redshift evolution of the SMD is a powerful probe of the nature of DM.  Integrating down to a limit of $\MUV =-16.5$ for the {\it James Webb Space Telescope (JWST)}, the SMD evolves as $\log$(SMD)$\propto -0.63 (1+z)$ in $\mx = 1.5\kev$ WDM, as compared to $\log$(SMD)$\propto -0.44 (1+z)$ in CDM. Thus high-redshift stellar assembly provides a powerful testbed for WDM models, accessible with the upcoming {\it JWST}.
\end{abstract}


\section{Introduction}
The current Lambda Cold Dark Matter ($\Lambda$CDM) cosmological paradigm has been extremely successful in explaining the distribution of matter from $\sim$Megaparsec (Mpc) to $\sim$Gigaparsec (Gpc) scales.  It reproduces the statistics of galaxy cluster abundances \citep[e.g][]{borgani-guzzo2001}, galaxy clustering on large scales \citep[e.g.][]{cole2005}, the temperature anisotropy of the cosmic microwave background \citep{lange2001,fixsen1996, hinshaw2013, planck2013} and the Lyman Alpha (Ly$\alpha$) forest \citep[e.g.][]{slosar2013}. In spite of this success, the validity of the $\Lambda$CDM paradigm has been a subject of close scrutiny given that it seems to predict too-much power on small scales: (i) CDM generically over-predicts the number of observed satellite and field galaxies \citep[e.g.][]{klypin1999,moore1999b, Papastergis11};
 (ii) CDM models predict halo profiles that are denser and cuspier than those inferred observationally \citep[e.g][]{navarro1997, subramanian2000}; and (iii) CDM predicts a population of massive, concentrated Galactic subhalos that are inconsistent with kinematic observations of Milky Way satellites \citep[e.g.][]{boylan2012}.

Baryonic feedback has had only moderate success in reconciling CDM with small-scale observations (e.g. \citealt{boylan2012, Garrison2013, Teyssier2013}).  A popular alternate solution can be found by appealing to the nature of dark matter itself. If dark matter consisted of $\sim$keV particles, so-called warm dark matter (WDM), it could smear out primordial small-scale density fluctuations, helping alleviate some tension with observations (e.g. \citealt{blumenthal1982, bode-ostriker2001, DeVega2012b}; though see e.g. \citealt{maccio2012_catch22}).  For thermal relics that decouple while relativistic, the free-streaming length can be easily translated to a particle mass, $\mx$. A number of different astrophysical approaches have already been used to obtain constraints on $\mx$. Using the Ly$\alpha$ forest power spectrum measured from high-resolution ($z>4$) quasar spectra \citet{viel2013} obtain a lower limit of $\mx \geq 3.3 \kev$. Using observations of dwarf spheroidal galaxies \citet{Devega2010} find $\mx > 1.0\kev$. \citet{kang2013} find that simultaneously reproducing stellar mass functions and the Tully-Fisher relation for $z=0-3.5$ galaxies requires $\mx \geq 0.75 \kev$. Using the number counts of high-$z$ gamma ray bursts (GRBs) with a highly-conservative Bayesian likelihood analysis, \citet{desouza2013} obtain limits of $\mx > 1.6-1.8 \kev$. Finally, using the number density of $z\approx10$ lensed galaxies and no astrophysical assumptions, \citet{pacucci2013} constrain $\mx \gsim 1\kev$. 

Given that structure formation proceeds hierarchically and WDM smears-out small-scale power, the effects of WDM would be manifested most strongly through a decrease in the number density and a change in the mass build-up of the smallest galaxies at the highest-$z$. This is an ideal time to use \highz ($z \gsim 7$) Lyman break Galaxies (LBGs) to probe the nature of WDM given that instruments like the Wide Field Camera 3 (WFC3) on the Hubble Space telescope (HST) have recently allowed a statistically significant sample of such galaxies to be collected. This sample has been used to obtain excellent constraints on the evolving ultraviolet luminosity function \citep[UV LF;][]{oesch2010, bouwens2010a,bouwens2011b, castellano2010, mclure2010, mclure2013, bowler2014, bradley2012, oesch2013, bouwens2014}. In addition, broad-band data (colours) have been used to infer the evolution of their global properties including the stellar mass density \citep[SMD;][]{gonzalez2011, stark2013,labbe2010a,labbe2010b}, specific star formation rates \citep[sSFR;][]{gonzalez2010, stark2009, yabe2009, schaerer2010, stark2013} and mass-to-light ratios \citep[M/L; Grazian et al., A\&A submitted; ][]{gonzalez2011}. 

In this work, we build a semi-analytic merger-tree based model that traces the growth of DM halos through mergers and smooth-accretion from the intergalactic medium (IGM). We include key baryonic processes of: {\it (i)} star formation and its impact on ejecting gas from the DM potential well through supernova (SN) winds (internal feedback), {\it (ii)} photo-heating of gas from the outskirts of galaxies exposed to an ultraviolet background (UVB) during reionization (external feedback), and {\it (iii)} the merger, accretion and ejection driven growth of stellar and gas mass. 
Using CDM as the baseline, we explore \highz galaxy formation for WDM masses of $\mx=1.5,3$ and 5 \kev. 

Our cosmological parameters are: ($\Omega_{\rm m },\Omega_{\Lambda}, \Omega_{\rm b}, h, n_s, \sigma_8) = (0.2725,0.702,0.04, 0.7, 0.96, 0.83)$, consistent with the latest results from the {\it Planck} collaboration \citep{planck20132}.  Unless stated otherwise, we quote all quantities in comoving units.

\section{Theoretical model }
\label{model}
We use the merger-tree based semi-analytic model presented in \citet{dayal2014} and apply it to four DM models: CDM, and WDM with $\mx =1.5,3$ and $5 \kev$. In brief, this model simultaneously traces the formation of successively larger DM halos through mergers of smaller progenitors, and follows their baryonic physics including star formation, the growth of stellar mass, and gas mass evolution due to accretion, ejection and mergers. Our model includes baryonic feedback from both SN explosions and photo-heating from reionization.  Below we elaborate on the main ingredients.

\subsection{DM merger trees}
\label{DM}

We construct CDM and WDM merger trees according to the prescription in \cite{benson2013} wherein the authors show that a number of modifications have to be introduced to obtain WDM merger trees and mass functions that are in agreement with N-body simulations. These include using a modified initial power spectrum that imposes a cut-off in power below a certain length scale depending on the WDM particle mass, using a critical over-density for collapse that depends on the WDM particle mass, using a sharp window function in {\it k}-space and calibrating the smooth-accretion of DM from N-body simulations. We start from 400 (800) $z=4$ galaxies for CDM and WDM with $\mx = 3$ and $5 \kev$ (WDM with $\mx = 1.5 \kev$), linearly distributed across the halo mass range $\log(M_h/\Msun) = 9-13$.  From these we build merger-trees using 320 equal redshift steps between $z=20$ and $4$ with a mass resolution of $M_{res} =10^8 \Msun$. As detailed in \citet{parkinson2008} and \citet{benson2013}, we use the modified binary merger tree algorithm with smooth accretion: we start with the conditional mass function given by the extended Press-Schechter theory (Bower 1991; Lacey \& Cole 1993),
\begin{eqnarray}
\lefteqn{\displaystyle{
f(M_1 \vert M_2)\, d\ln M_1 = 
\sqrt{\frac{2}{\pi}} \,
\frac{\sigma_1^2
  (\delta_1-\delta_2)}{[\sigma_1^2-\sigma_2^2]^{3/2}} } \, \times} &&
\nonumber \\
&&\displaystyle{ \exp\left[ - \frac{1}{2} \frac{(\delta_1-\delta_2)^2}{(\sigma_1^2-\sigma_2^2)}\right]
\left\vert \frac{d\ln\sigma}{d\ln M_1} \right\vert \,
d\ln M_1}.
\end{eqnarray}
Here $f(M_1 \vert M_2)$ represents the fraction of mass from halos of mass $M_2$ at redshift $z_2$ that is contained in progenitor halos of mass $M_1$ at an earlier redshift $z_1$, while $\delta_1$ and $\delta_2$ are the linear density thresholds for collapse at these two redshifts. The rms linear density fluctuation extrapolated to $z=0$ in spheres containing mass $M$ is denoted $\sigma(M)$ with $\sigma_1\equiv\sigma(M_1)$ and $\sigma_2\equiv\sigma(M_2)$.
As a consequence, the mean number of halos of mass $M_1$ into which a halo of mass $M_2$ splits when one takes a step $dz_1$ up in redshift is:
\begin{equation}
\frac{dN}{d M_1} =  \frac{1}{M_1} \ \frac{df}{dz_1} \frac{M_2}{M_1} dz_1
\qquad (M_1 < M_2) .
\label{eq:mean_density}
\end{equation}
In order to reduce the systematic differences between merger trees predictions and N-body simulations, Parkinson et al. (2008) introduced a slight modification to Eqn. \ref{eq:mean_density}, by making the following substitution:
\begin{equation}
  \frac{dN}{dM_1} \rightarrow  \frac{dN}{dM_1} \ 
G(\sigma_1/\sigma_2,\delta_2/\sigma_2) .
\label{eq:mean_density_mod}
\end{equation}
With the assumption that:
\begin{equation}
G(\sigma_1/\sigma_2,\delta_2/\sigma_2) = G_0  \
\left( \frac{\sigma_1}{\sigma_2} \right)^{\gamma_1} \
\left( \frac{\delta_2}{\sigma_2} \right)^{\gamma_2} ,
\end{equation}

Then, by specifying a required mass resolution, $M_{\rm res}$, for the algorithm one can integrate to determine the mean number of progenitors with masses $M_1$ in the interval $M_{\rm res}<M_1<M_2/2$:
\begin{equation}
P=\int_{M_{\rm res}}^{M_2/2} \frac{dN}{d M_1} \ dM_1,
\end{equation}
and the fraction of mass of the final object in progenitors below this resolution limit:
\begin{equation}
F= \int_{0}^{M_{\rm res}} \frac{dN}{d M_1}\ \frac{M_1}{M_2} \
dM_1,
\end{equation}

At a given step time-step, a halo could have progenitors above or below the resolution limit, $M_{res}$. The masses of progenitors below the resolution limit are treated as `smooth-accretion' from the IGM.
We scale the relative abundances of the merger tree roots to match the $z=4$ Sheth-Tormen halo mass function \citep{sheth-tormen1999}.

\subsection{External feedback from reionization} 
\label{uv_fb}
Cosmic reionization suppresses the baryonic content of galaxies by photo-heating gas at their outskirts \citep{klypin1999, moore1999, somerville2002} thereby preventing efficient cooling: what we refer to below as ``external feedback''. External feedback depends on the local thermal histories of gas near halos.  As reionization is poorly constrained and very inhomogeneous, this feedback is difficult to quantify. Recently \citet{sobacchi2013a} have run a large suite of 1d cosmological collapse simulations to explore the parameter space allowed by the inhomogeneity of reionization. These authors provide a functional form, motivated by linear theory, for the critical halo mass ($M_{crit}$) that can retain half of its baryons compared to the global value at any redshift $z$:

\begin{equation}
M_{crit}(z)= M_0 J_{21}^a \bigg(\frac{1+z}{10}\bigg)^b \bigg[1 - \bigg(\frac{1+z}{1+z_{IN}}\bigg)^c\bigg]^d,
\end{equation}
where $J_{21}$ represents the ionizing UVB intensity in units of $10^{-21} {\rm erg\, s^{-1}\, Hz^{-1} \, cm^{-2} \, sr^{-1}}$, $z_{IN}$ is the redshift at which the  halo is exposed to the UVB (we take $z_{IN}=9$ for this work), and the best-fit parameters are $(M_0,a,b,c,d,J_{21}) = (2.8\times 10^9 \Msun, 0.17, -2.1,2.0,2.5,0.01)$. \citet{sobacchi2013a} also show that the baryon fraction (with respect to the cosmic mean) of a halo of mass $M_h$ is well-fit with the simple form: 
\begin{equation}
f_b(z)= 2^{-M_{crit}(z)/M_h(z)},
\end{equation}
Here we use the same parameter values for external feedback for all the four DM models explored in this paper: CDM and WDM with $m_x = 1.5,3$ and $5$ \kev.

\subsection{Internal feedback from supernovae} 
\label{sn_fb}
We also include the effects of internal feedback, i.e. gas ejection from DM halos by SN driven winds. As detailed in \citet{dayal2014}, we build a model based on the following simple idea: the SN kinetic energy from star formation exceeding the binding energy of a halo will result in a complete loss of gas mass, quenching further star formation (``inefficient star-formers"). However, halos where the binding energy exceeds the SN kinetic energy will only lose part of their gas and  continue forming stars (``efficient star-formers"). 

\begin{figure*} 
   \center{\includegraphics[scale=0.82]{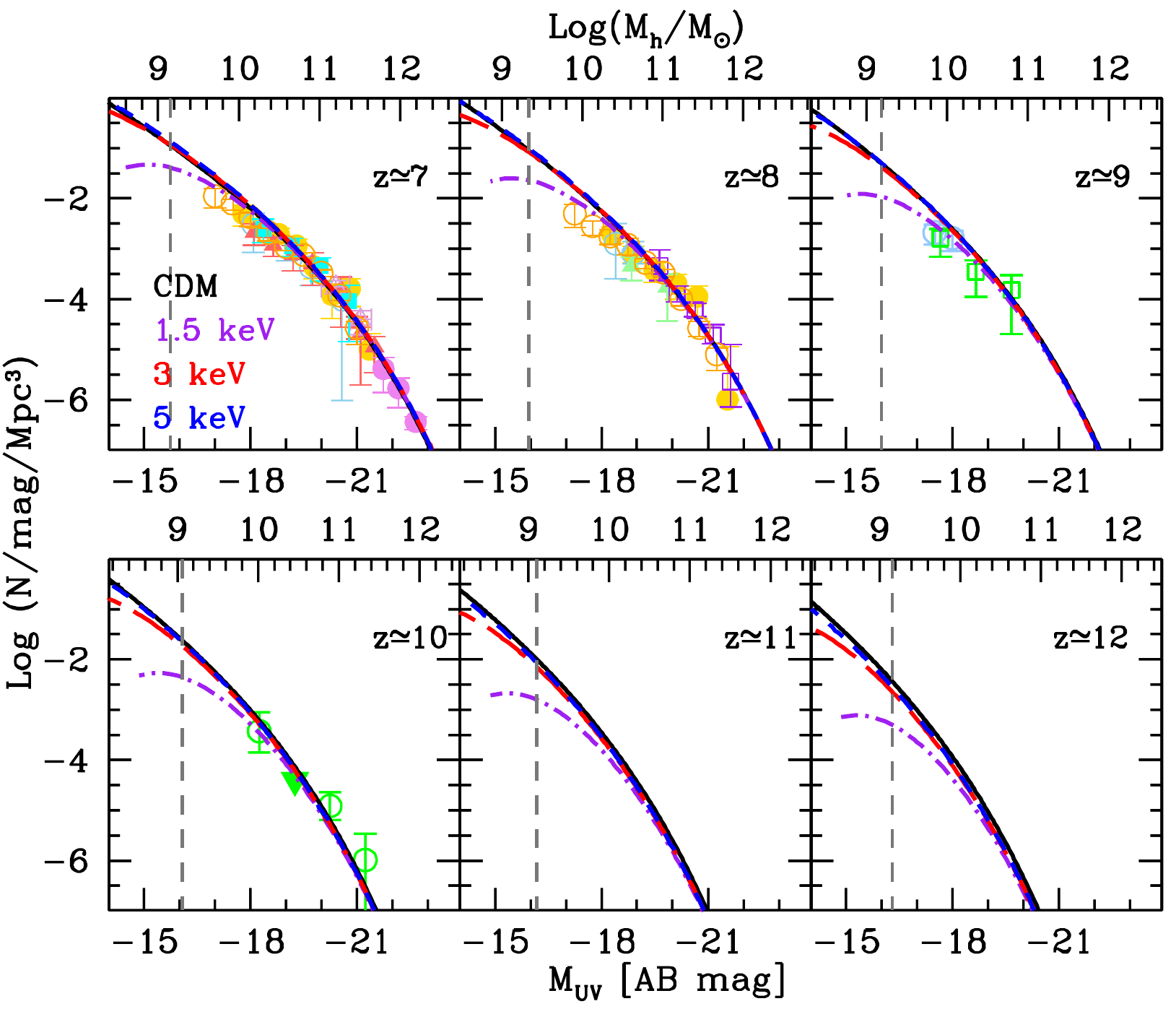} }
  \caption{The evolving LBG UV LFs at $z \simeq 7-12$ for the four different DM models considered, obtained by scaling the appropriate HMF with a halo mass independent star formation efficiency of $f_* = 0.9\%$ at $z \simeq 7$ and $f_* = 1.3\%$ for $z \gsim 8$.
 In all panels, lines show theoretical results for CDM (black solid line) and WDM with particle masses of $5 \kev$ (blue short-dashed), $3 \kev$ (red long-dashed) and $1.5\kev$ (violet dot-dashed); dashed vertical lines show the $10\sigma$ $10^4 \, {\rm s}$ integration limits of the JWST. In all panels points show observational results: (a) $z \simeq 7$: \citet[filled cyan squares]{oesch2010}, \citet[empty blue circles]{bouwens2010a}, \citet[filled yellow circles]{bouwens2011b}, \citet[empty purple triangles]{castellano2010}, \citet[filled red triangles]{mclure2010}, \citet[empty orange circles]{mclure2013} and \citet[filled purple circles]{bowler2014}; (b) $z \simeq 8$: \citet[empty blue circles]{bouwens2010a}, \citet[filled yellow circles]{bouwens2011b}, \citet[filled green triangles]{mclure2010}, \citet[empty purple squares]{bradley2012} and \citet[empty orange circles]{mclure2013}, (c) $z \simeq 9$: \citet[empty blue circles]{mclure2013} and \citet[empty green squares]{oesch2013} and, (d) $z \simeq 10$: \citet[empty green circles]{bouwens2014}; the downward pointing triangle represents the upper-limit of the $z\simeq 10$ data at $\MUV \simeq -19.25$. }
\label{scaled_uvlf} 
\end{figure*}

To implement this idea, we start by calculating the {\it ejection efficiency} ($f_*^{ej}$) which is the the fraction of gas that must be converted into stars so that the SN ejection energy ($E_{SN}$) equals the binding energy ($E_{ej}$).
We take 
\begin{equation}
E_{SN}= f_w \nu E_{51} \M*(z),
\end{equation}
where $\M*(z)$ is the newly formed stellar mass, $E_{51} = 10^{51} {\rm ergs}$ is the ejection energy per SN, $\nu = (134\Msun)^{-1}$ is the fraction of stars that explode as SN for the chosen Salpeter IMF between $0.1-100\Msun$ and $f_w$ is the fraction of total SN energy that drives winds. Further, 
\begin{equation}
E_{ej} = \frac{1}{2} [M_{g,i}(z)-\M*(z)] v_e^2,
\end{equation}
where $M_{g,i}(z)-\M*(z)$ implies that SN explosions have to eject the part of the initial gas mass not converted into stars, and the escape velocity $v_e$ can be expressed in terms of the halo rotational velocity ($v_c$) as $v_e = \sqrt 2 v_c$. The {\it effective efficiency} of star formation can then be expressed as $f_*^{eff} =min[f_*,f_*^{ej}]$: while galaxies hosted in large DM halos (efficient star formers) can continuously convert a fraction $f_*$ of their gas into stars, smaller halos (feedback-limited systems/inefficient star formers) can form stars with a maximum efficiency $f_*^{ej}$ that decreases with decreasing halo mass. Note that this formalism maximizes the impact of internal feedback by limiting star formation in small-mass halos to be just sufficient to evacuate the remaining gas from their halos. Therefore, below we also present results (Fig. \ref{scaled_uvlf}) which ignore internal feedback entirely, thus bracketing the expected astrophysical uncertainties.  In any case, our results are driven by the {\it relative} difference between CDM and WDM, which is more robust to the astrophysical details.

\subsection{Implementing baryonic physics into merger trees} 
\label{sec_mt}
We now briefly describe how star formation and feedback prescriptions are implemented into the DM merger trees. The formalism below closely follows that presented in \citet{dayal2014}, to which we refer the interested reader for more details. 

Once the merger tree for each galaxy has been constructed, we proceed forward in time from the highest merger tree output redshift, $z=20$. Starting from the first DM progenitor (with halo mass $M_0$) along a branch of the merger tree, we assign to it an initial gas mass $M_{g,i} (z) = f_b(z) (\Omega_b/\Omega_m) M_0 (z)$, where $f_b(z)$ is the baryon fraction discussed in Sec. \ref{uv_fb}. A part of this gas mass gets converted into newly formed stellar mass $\M*(z)$, 
\begin{equation}
M_*(z) = f_*^{eff} M_{g,i}(z) \frac{\Delta t}{t_{sf}},
\label{mstarnew}
\end{equation}
where $\Delta t$ is the time difference between the merger tree output at $z$ and the successive step at $z -\Delta z$ where $\Delta z =0.05$. Further, $t_{sf}$ is the star formation timescale which we take to scale with the Hubble time ($t_H$) at the redshift considered such that $t_{sf} \approx 0.01 t_H(z)$.

 To maintain simplicity, we assume that every new stellar population has a fixed metallicity of $0.05 \Zsun$ and an age of $2\, {\rm Myr}$ so that its UV luminosity (at $\lambda = 1500$\,\AA) can be calculated as $L_{UV} = 10^{33.077} (\M*/\Msun)\, {\rm erg \, s^{-1} \AA^{-1}}$ using the population synthesis code {\small STARBURST99} \citep{leitherer1999}.

The formation of this stellar mass results in $M_{g,ej}(z)$ of gas mass being ejected at the given redshift step:
\begin{equation}
M_{g,ej}(z) = [M_{g,i}(z) - M_{*}(z)] \frac{f_*^{eff}}{f_*^{ej}},
\label{mej}
\end{equation}
where some of the initial gas mass has been converted into stars. The final gas mass, $M_{g,f}(z)$, that remains in the galaxy at the end of that redshift-step is then
\begin{equation}
M_{g,f}(z) = [M_{g,i}(z) - \M*(z)] \bigg[1-\frac{f_*^{eff}}{f_*^{ej}}\bigg].
\label{mgf}
\end{equation}
As seen from Eqn. \ref{mgf}, the final gas mass depends on the ratio of $f_*^{eff}$ and $f_*^{ej}$: galaxies which form stars at an efficiency capable of ejecting the rest of the gas (i.e. $f_*^{eff} = f_*^{ej}$) will lose all their gas mass and undergo dry mergers, contributing only stellar mass to their successor. However, galaxies forming stars at a lower efficiency than that required to eject the rest of the gas ($f_*^{eff}<f_*^{ej}$) will only lose a part of their gas content, resulting in wet mergers, bringing in both gas and stellar mass.  
 
On the other hand, a galaxy that has progenitors inherits a certain amount of stars and gas from them following merging events. In addition, this galaxy also obtains a part of its DM (and gas) 
mass through smooth-accretion from the IGM: while in principle a cosmological ratio of DM and baryons can be accreted onto the halo, UVB photo-heating feedback suppresses the available gas reservoir for accretion inside the ionized IGM as explained in Sec. \ref{uv_fb}. Thus, the total initial gas mass in the galaxy at $z$ is the sum of the 
newly accreted gas mass, as well as that brought in by its merging progenitors.

This updated gas mass is then used to calculate the new stellar mass formed in the galaxy 
as described by Eqn. \ref{mstarnew}. The total stellar mass in this galaxy is now the sum 
of mass of the newly-formed stars, and that brought in by its progenitors.

\begin{figure*} 
   \center{\includegraphics[scale=0.85]{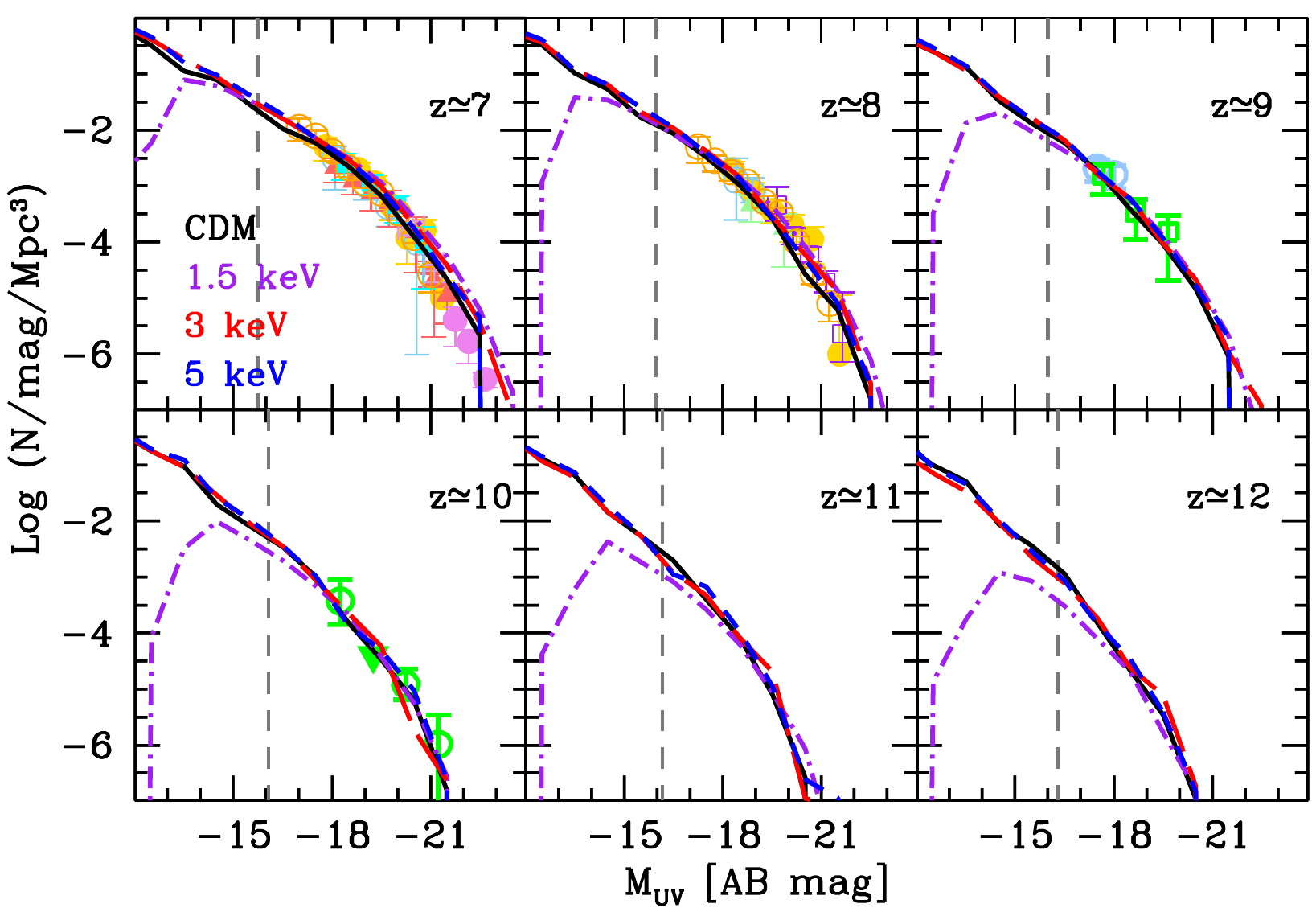} }
  \caption{The evolving LBG UV LF at $z \simeq 7-12$ in different DM models, computed with our fiducial semi-analytical galaxy formation model. In all panels, lines show the results using the fiducial model that invokes a total of two redshift and mass-independent free parameters: the star formation efficiency ($f_*\approx0.04$) and the fraction of SN energy driving winds ($f_w=0.1$). In all panels lines show theoretical results for CDM (black solid line) and WDM with particle masses of $5 \kev$ (blue short-dashed), $3 \kev$ (red long-dashed) and $1.5\kev$ (violet dot-dashed); dashed vertical lines show the $10\sigma$ $10^4 \, {\rm s}$ integration limits of the JWST. In all panels, points show observational results (see Fig. \ref{scaled_uvlf} for references).}
\label{fiducial_uvlf} 
\end{figure*}

Our fiducial parameters are selected to match the observed UV LF.  Specifically, we take $f_* = 0.038$ and $f_w = 0.1$ which result in a good fit to available data at $z \simeq 7-10$ for all the four DM models considered (see Fig. \ref{fiducial_uvlf}).  Roughly speaking, $f_w$ affects the faint-end slope of the UV LF where feedback is most effective, while $f_*$ determines the amplitude and normalization at the bright-end where galaxies can form stars with the maximum efficiency. Although this model need not be unique in describing the observed LF, we stress again that our main conclusions are driven by the relative differences between the cosmologies, which are more robust to astrophysical uncertainties.

\section{Early galaxy evolution in different dark matter models}
We now show how \highz galaxy assembly varies with the DM particle mass considered, and its impact on observables including the UV LF, the M/L relation and the SMD. 

\subsection{Ultraviolet luminosity functions}
\label{uvlf}
The evolving UV LF is the most robust piece of information available for $z \gsim 7$ galaxies, with the observational estimates for a number of different groups \citep[e.g.][]{oesch2010, bouwens2010a,bouwens2011b, castellano2010, mclure2010, mclure2013, bowler2014, bradley2012, oesch2013, bouwens2014} being in good agreement. The simplest approach to obtaining a UV LF is to scale the halo mass function (HMF) at that redshift assuming a fixed halo mass-to-light ratio. Indeed, \citet{schultz2014} propose that the cumulative number density of high-redshift galaxies could be used to constrain $\mx$. We caution however that constraints on WDM obtained through such abundance matching directly rely on the assumed halo mass $\leftrightarrow$ UV luminosity relation, which \citet{schultz2014} take to be independent of the DM model and assume a power-law extrapolation towards small masses.  As we shall see below, this need not be the case.

Before presenting results from our complete model which includes feedback, in Fig. \ref{scaled_uvlf} we show UV LFs obtained simply by multiplying the HMFs by a constant mass-to-light ratio. Matching to the observations requires a halo star formation efficiency with values $f_* = (0.9,1.3)\%$ for $z=(7,8-10)$; we use $f_* = 0.013$ for all $z \simeq 11$ and $12$ given the lack of data at these $z$. A constant halo mass $\leftrightarrow$ UV luminosity mapping allows us to estimate which halos host observable galaxies: e.g. galaxies with $\MUV \simeq -15\, (-20)$ reside in halos with $M_h \simeq 10^{8.6-8.8}\, (10^{10.8-11.2}) \Msun$ at $z=7-12$.

\begin{table} 
\begin{center} 
\caption {For the redshift shown in column 1, we show the observed faint-end slope of the UV LF \citep{mclure2009, mclure2013} in column 2. Columns 3 and 4 show the faint-end slope of the fiducial UV LFs with the $1-\sigma$ errors for CDM and $1.5\kev$ WDM, respectively. The faint-end slopes for the theoretical UV LF have been computed over the absolute magnitude range $-18 \leq M_{UV} \leq -14$.} 
\begin{tabular}{|c|c|c|c|c}
\hline 
$z$& $\alpha_{obs}$ & $\alpha_{\rm CDM}$ & $\alpha_{1.5\kev}$  \\  
\hline 
$7$& $-1.90^{+0.14}_{-0.15}$ & $-1.96 \pm 0.18$  & $-1.85 \pm 0.11$\\ 
$8$& $-2.02^{+0.22}_{-0.23}$ & $-2.06 \pm 0.22$  & $-1.93 \pm 0.13$  \\ 
$9$& $-$ & $-2.21 \pm 0.32$ & $-2.01 \pm 0.16$\\ 
$10$& $-$ & $-2.31 \pm 0.45$ & $-2.10 \pm 0.18$\\ 
$11$& $-$ &  $-2.39 \pm 0.32$ & $-2.22 \pm 0.28$\\ 
$12$& $-$ &  $-2.62\pm 0.53$ & $-2.34 \pm 0.44$\\ 
\hline
\label{table1} 
\end{tabular} 
\end{center}
\end{table}

\begin{figure*} 
  \center{\includegraphics[scale=0.75]{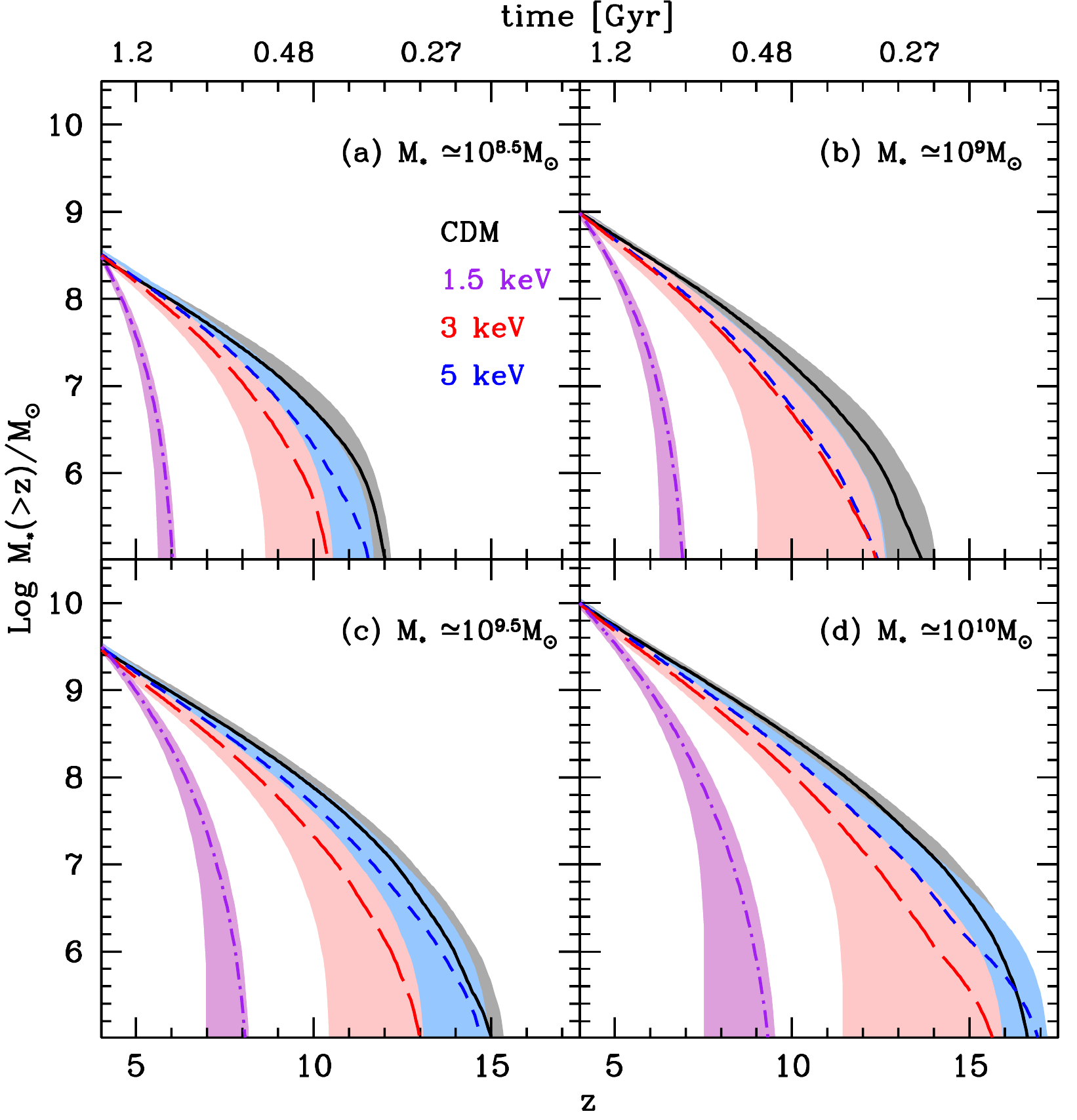}}
  \caption{Average stellar mass assembly of galaxies as a function of $z$. For the final $z=4$ $\M*$ value quoted in each panel, we show the stellar mass build up for the four different DM models considered: CDM (solid black line), $5 \kev$ WDM (blue short-dashed line), $3 \kev$ WDM (red long-dashed line) and $1.5 \kev$ WDM (violet dot-dashed line). Gray, blue, red and purple shaded regions show the $1-\sigma$ dispersion for the CDM and WDM models of mass $5,3$ and $1.5\kev$, respectively. As seen, high-$z$ star formation is more rapid in WDM models. For example, $z=4$ galaxies with $\M*=10^{8.5} \Msun$ assemble 90\% of their stellar masses within the previous 1.03 (0.64) Gyr in CDM (1.5 keV WDM). This younger stellar population means that for a given stellar mass, WDM galaxies are more UV luminous. }
\label{fig_assembly} 
\end{figure*}

From Fig. \ref{scaled_uvlf} we can also estimate the viability of distinguishing between different DM models. We see that the WDM LFs with $\mx = 3 \, (5) \kev$ are essentially indistinguishable from CDM down to an absolute magnitude of $\MUV \simeq -15\, (-14)$; this is about 0.5 (1.5) magnitudes fainter than the range of the next generation of instruments such as the JWST. However, the UV LF for WDM with $\mx = 1.5 \kev$ starts to ``peel-away" from the CDM UV LF at a value of $M_h \simeq 10^{10} \Msun$ at all $z =7$ to 12, and exhibits faint-end slope values shallower than the value of $\alpha \simeq -2$ inferred observationally \citep[e.g.][]{mclure2013}. In spite of this peel-away, the faint-end slope values for all the four DM models explored here are equally compatible with current observations, including the deepest $z \simeq 7,8$ data obtained from the Hubble Ultra Deep Field 2012 \citep[HUDF12;][]{ellis2013, mclure2013}. With its higher sensitivity \footnote{We use the detection limits for a $10\sigma$ $10^4 \, {\rm s}$ observation provided at http://www.stsci.edu/jwst/instruments/nircam/sensitivity/table.}, the {\it JWST} could potentially constrain the UV LF and hence $\alpha$ to fainter magnitudes, allowing constraints on $\mx$. 

However, the UV LFs are more complex, shaped by the star-formation histories of each galaxy. While galaxies at the faint end of the UV LF build up most of their gas mass (and hence luminosity) by smoothly-accreting gas from the IGM due their tiny progenitors being SN feedback-limited, the gas mass build-up for galaxies at the bright end is dominated by mergers of gas-rich progenitors. Therefore, for the remainder of this paper, we use our more physical model for galaxy evolution, described in Sec. \ref{model}.  The corresponding UV LFs are shown in Fig. \ref{fiducial_uvlf}, showing similar trends already noted for the simple HMF scalings in Fig. \ref{scaled_uvlf}. While our CDM and $\mx =3,5 \kev$ WDM models are consistent down to $\MUV \simeq -12$ for all $z \simeq 7-12$, the $1.5$ keV model shows a dearth of galaxies fainter than the JWST limit of $M_{UV} =-16$ at $z\geq10$. Although our fiducial model results are also in reasonable agreement with observations at $z \simeq 5,6$, we focus at $z \gsim 7$ in this paper since the differences between CDM and WDM are expected to become increasingly pronounced with increasing $z$.


\begin{figure*} 
   \center{\includegraphics[scale=0.85]{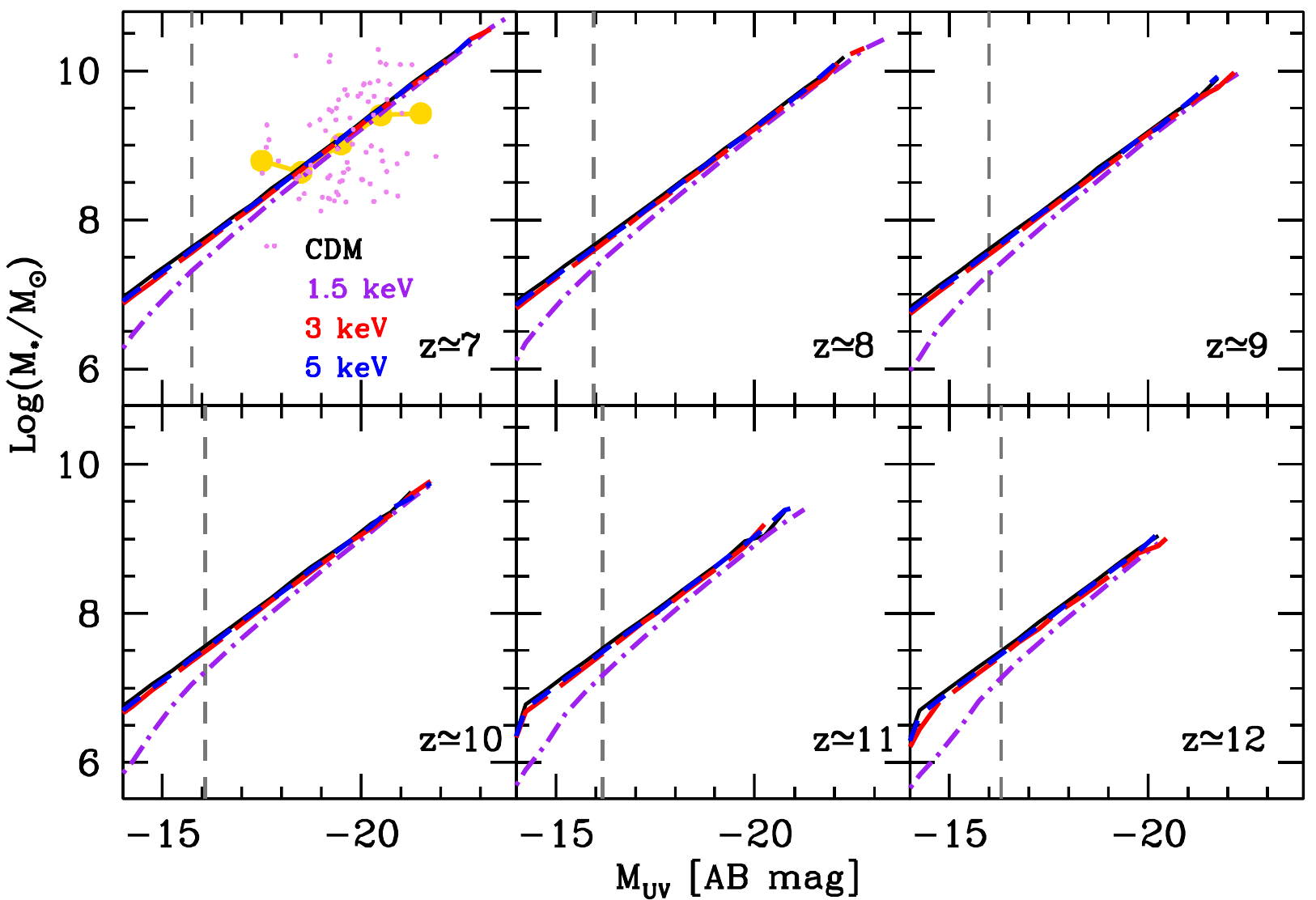} }
  \caption{Mass-to-light relation showing galaxy stellar mass as a function of UV magnitude for $z \simeq 7-12$ as marked. In each panel we show average $\M*$ values for given $\MUV$ bins from our fiducial model for the four DM models considered in this work: CDM (solid black line), $5 \kev$ WDM (blue short-dashed line), $3 \kev$ WDM (red long-dashed line) and $1.5 \kev$ WDM (violet dot-dashed line). At $z \simeq 7$, violet points show the values for real galaxies 
in the CANDELS and HUDF fields, with yellow points showing the observed medians in each UV bin as inferred by Grazian et al. (A\&A submitted). In each panel, dashed vertical lines show the $10\sigma$ $10^4 \, {\rm s}$ integration limits of the JWST.}
\label{fig_m2l} 
\end{figure*}

Interestingly, our fiducial model based on merger histories and feedback {\it decreases} the differences between WDM and CDM LFs (see Fig. \ref{fiducial_uvlf}), compared to the simple HMF scaling shown in Fig. \ref{scaled_uvlf}.  This is due to the fact that stellar populations are on average younger in WDM, and so are more UV luminous (for a given stellar mass).  Thus the dearth of small-mass halos in WDM is somewhat countered by their lower mass-to-light relation.  We elaborate further on this in Sec. \ref{assembly}.

To summarise, our fiducial model shows that the evolving LBG UV LFs in WDM models with masses of $\mx \geq 3\kev$ are indistinguishable from CDM down to $\MUV =-12$ for $z \simeq 7-12$. However, the LBG UV LF in the $1.5 \kev$ WDM scenario shows a shallower (by about 0.1-0.3) faint-end slope ($\alpha$) compared to the three other models at all $z \simeq 7-12$ as shown in Table \ref{table1}; this naturally leads to a decrease (of about 0.5 dex) in the number density of faint sources ($\MUV \simeq -15,-16$) at $z=11,12$. This slope difference is too small to be distinguishable with current data where $\alpha$ and the knee of the UV LF ($M_{UV,*}$) are degenerate \citep[e.g.][]{mclure2013, bouwens2014}. Even with its sensitivity extending down to $\MUV \simeq -16.5$ at these $z$, it is doubtful whether the {\it JWST} will be able to accurately pin down the shape and the faint-end slope value of the UV LF and allow differentiating between WDM models with $\mx \lsim 2 \kev$ and $\mx \gsim 2 \kev$.

\subsection{Assembling early galaxies}
\label{assembly}
We now explore the build-up of the LFs shown in the previous section.  Due to the suppression of small-scale structure in WDM models, star-formation is delayed and more rapid \citep[e.g][]{calura2014, sitwell2014}.  We quantify this for our galaxy evolution models in Fig. \ref{fig_assembly}, in which we show the stellar mass assembly histories for four different mass bins ranging from $\M* = 10^{8.5}-10^{10} \Msun$.

As expected in hierarchical structure formation, the larger the final stellar mass, the earlier it started forming (i.e. with flatter assembly histories).  For example, $z=4$ galaxies with $\M* = 10^{10} \Msun$ build up 90\% of their stellar mass within the last 1.26 Gyr in CDM.  Smaller galaxies with $\M* = 10^{8.5} \Msun$ in CDM take only 1.03 Gyr to build-up 90\% of the stellar mass. This distinction is even more dramatic in WDM models.  For example, $z=4$ galaxies with $\M*=10^{8.5} \, (10^{10}) \Msun$ assemble 90\% of their stellar masses within the previous 0.64 (1.12) Gyr, for $m_x=$1.5 keV.  The distinction of WDM with respect to CDM is most notable in the dearth of small halos, near the atomic cooling threshold (e.g. Fig. 1). The lack of these progenitor building blocks results in a sudden appearance of galaxies in WDM models, with little scatter in the assembly history.

As shown in Sec. \ref{uvlf}, the higher particle mass WDM models are difficult to distinguish from CDM with the assembly histories for $\mx \geq 3\kev$ WDM models only differing from CDM in the \highz tails. Indeed, stellar mass assembly histories in CDM and the $5\kev$ WDM model differ by less than 50 Myr, throughout the range shown in Fig. \ref{fig_assembly}.


\subsection{Mass to light relation}
\label{sec_m2l}
 
\begin{figure*} 
   \center{\includegraphics[scale=0.85]{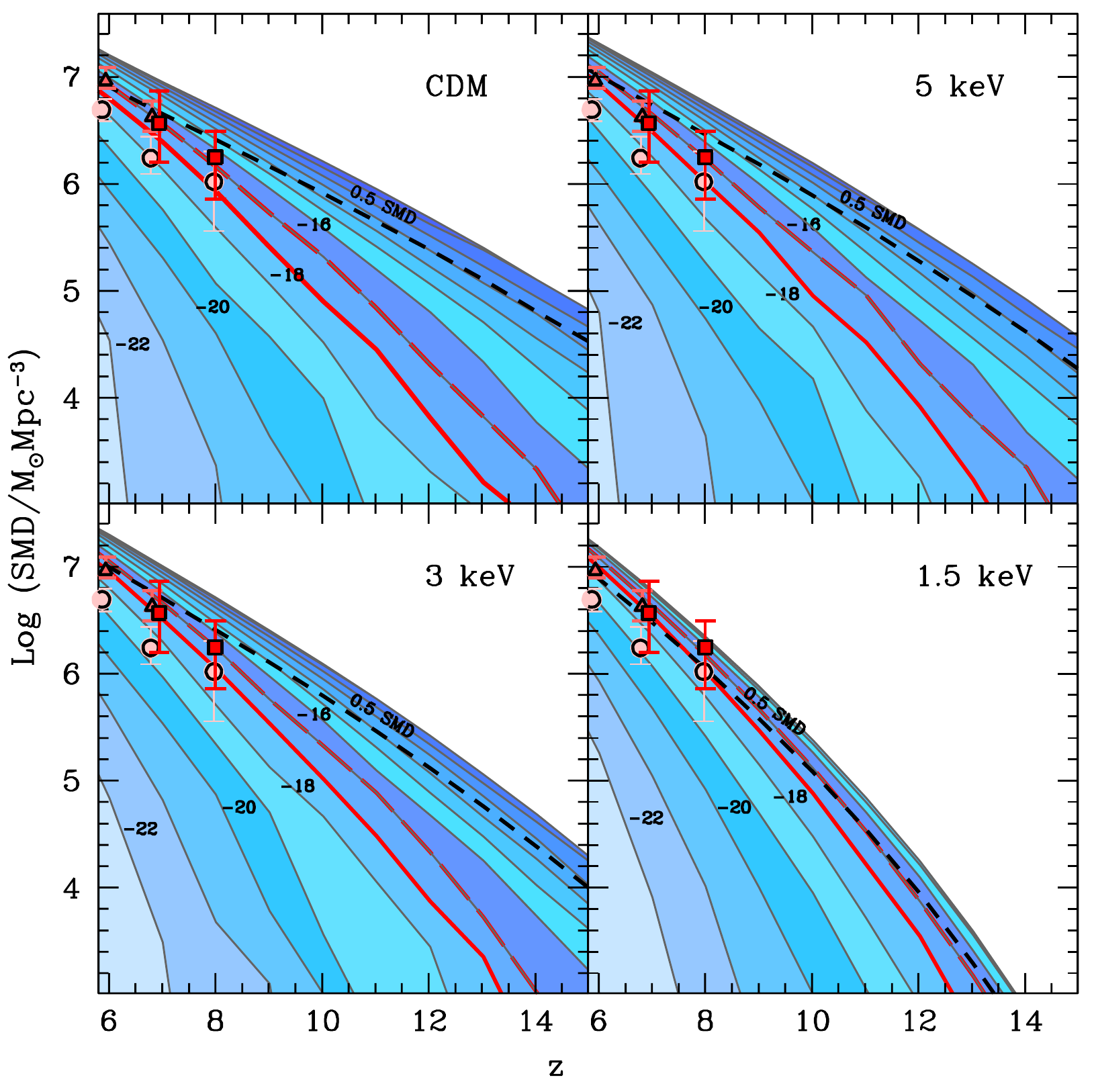} }
  \caption{The SMD as a function of redshift for the four DM models considered in this work (CDM and WDM of $\mx =1.5,3,5\kev$), as marked. The different coloured contours show the contribution to the total SMD from galaxies brighter than the magnitude value marked in the contour (we only mark every alternate contour for clarity). In each panel the solid red line shows the SMD from galaxies that have already been detected ($\MUV \leq -18$) to allow comparison with the data points: \citet[filled triangles]{gonzalez2011}, \citet[filled circles]{stark2013} and \citet[filled squares]{labbe2010a,labbe2010b}. In each panel, the short-dashed black line shows the value of 50\% of the total SMD at any $z$ for the appropriate DM model; the long-dashed red line in each panel shows the CDM SMD integrating down to a very conservative magnitude limit of $M_{UV} = -17$ for the JWST. } 
\label{fig_smd} 
\end{figure*}

In the previous section we saw that galaxies in WDM models assemble their stars more rapidly compared to CDM.  This rapid assembly translates to a younger, more UV luminous stellar population.  A useful observational probe of this trend is the mass to light relation, which links the total stellar mass ($\M*$) and the UV magnitude ($\MUV$).

In Fig. \ref{fig_m2l} we show the $\M*-\MUV$ relation for our DM models. The $\M*-\MUV$ relation for CDM (and $\mx \geq 3\kev$ WDM) galaxies brighter than $M_{UV}=-15$ is well fit by a power law:
\begin{equation}
\label{eq_m2l}
{\rm log}\, \M* = \beta \MUV + \gamma,
\end{equation}
where $\beta = -0.38$ and $\gamma = 2.4-0.1z$. This relation is in good agreement both with estimates using abundance matching (e.g. \citealt{KF-G12, schultz2014}) and direct observational estimates for LBGs (Grazian et al., A\&A submitted); we show the last group's results in the $z =7$ panel who also find a slope of $\beta = -0.4$.

As seen from Eqn. \ref{eq_m2l}, the normalisation ($\gamma$) of the $\M*-\MUV$ relation decreases with increasing $z$ (although the slope remains unchanged), i.e. a given UV luminosity is produced by lower $\M*$ galaxies with increasing $z$.  This is due to the fact these galaxies are hosted by increasingly (with $z$) rare, biased halos, farther on the exponential tail of the HMF, whose fractional growth is more rapid.

While there is little difference between the $\M*-\MUV$ relation for CDM and $\mx \geq 3 \kev$ WDM, this relation starts diverging from the CDM one at $\MUV \simeq -19 $ at all $z=7-12$ in the $1.5\kev$ WDM model.  The relation for $M_{UV} \leq -15$ for $1.5\kev$ is well-fit by Eqn. \ref{eq_m2l} with $\beta = -0.01z -0.34$ and $\gamma = -0.31z+2.93$.
This steeper $\M*-\MUV$ relation implies that galaxies with a given stellar mass {\it are more UV luminous} in WDM cosmologies.  This trend is driven by the more rapid assembly, and associated younger, more UV luminous stellar population, as we have seen in Fig. \ref{fig_assembly}.

Hence, a  steepening of the $\M*-\MUV$ relation towards faint ($\MUV \gsim -19$) galaxies is evidence of WDM with $\mx \lsim 2$ keV. Using this relation to differentiate between DM models relies on accurate estimates of $\M*$ values. Hence it will only be practical for relevant $\mx$ values with the advent of {\it JWST} and its improved near-infrared (NIR) data.  We also caution that self-consistently estimating $\M*$ from multi-band photometry should use the correct star-formation histories corresponding to each DM model.

\begin{figure*} 
   \center{\includegraphics[scale=0.85]{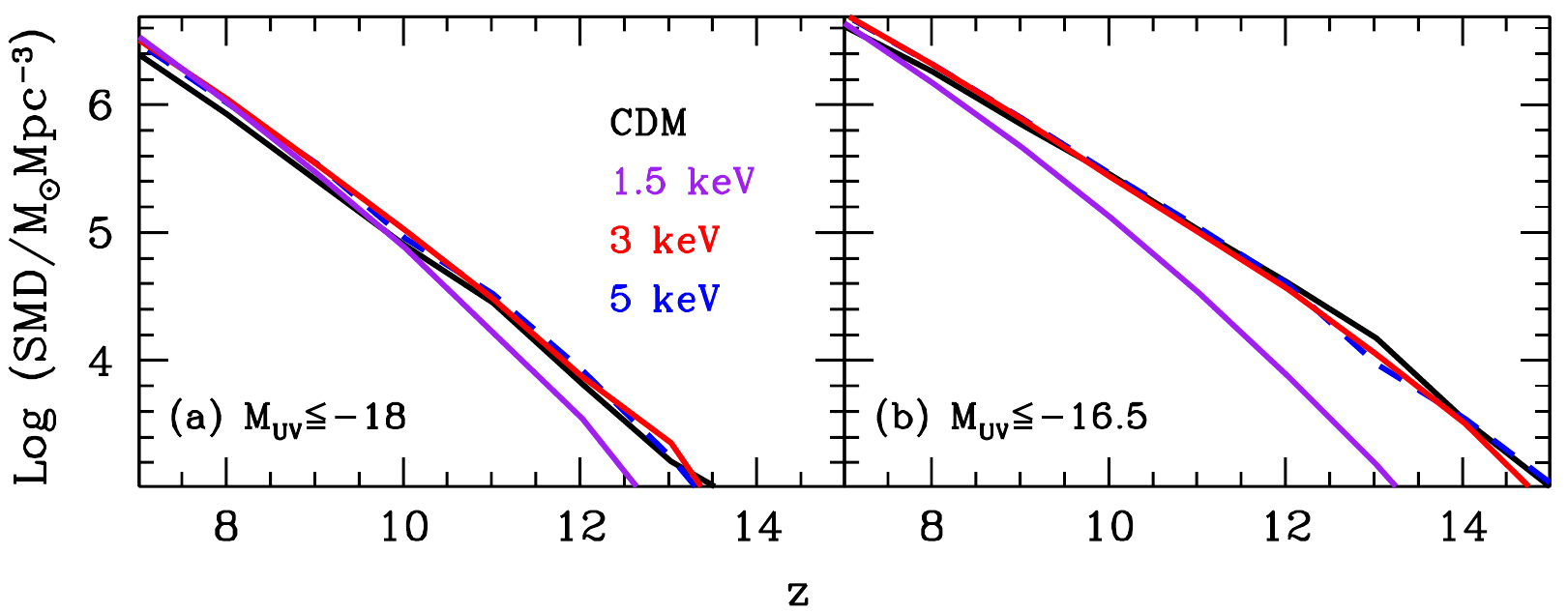} }
  \caption{Redshift evolution of the SMD for the four DM models considered in this work: CDM (black line), $5 \kev$ WDM (blue line), $3 \kev$ WDM (red line) and $1.5 \kev$ WDM (violet line). The left and right panels show the SMD from galaxies that have already been detected, i.e. $\MUV \leq -18$, and galaxies that are expected to be detected using a magnitude limit of $\MUV \leq -16.5$ for the JWST, respectively. As seen, while the SMD measured by the JWST will be indistinguishable for CDM and WDM with $\mx =3$ and $5 \kev$, our model predicts that these three models will have formed about 3 (10) times more stellar mass per unit volume at $z \simeq 11$ ($z \simeq 13$) compared to the $1.5\kev$ case, providing one of the strongest hints on the nature of DM using \highz galaxies. } 
\label{fig_smdpred} 
\end{figure*}


\subsection{Stellar mass density (SMD) evolution with z}
We now combine the trends noted above, showing predictions of the redshift evolution of the SMD. Integrated down to a given UV sensitivity threshold, it provides a straightforward estimate of how well upcoming observations can discriminate amongst DM models.  In Fig. \ref{fig_smd}, we show that the assembly of galaxies is very similar in CDM compared to the $3$ and $5 \kev$ WDM models leading to their SMD contributions from different $\MUV$ bins being almost identical: while observed galaxies ($\MUV \leq -18$) make up 50\% of the total SMD at $z=6$, galaxies fainter by about 1.5 orders of magnitude ($\MUV\leq -14)$ make up 50\% of the SMD at $z \simeq 9$.

\begin{figure*} 
   \center{\includegraphics[scale=0.85]{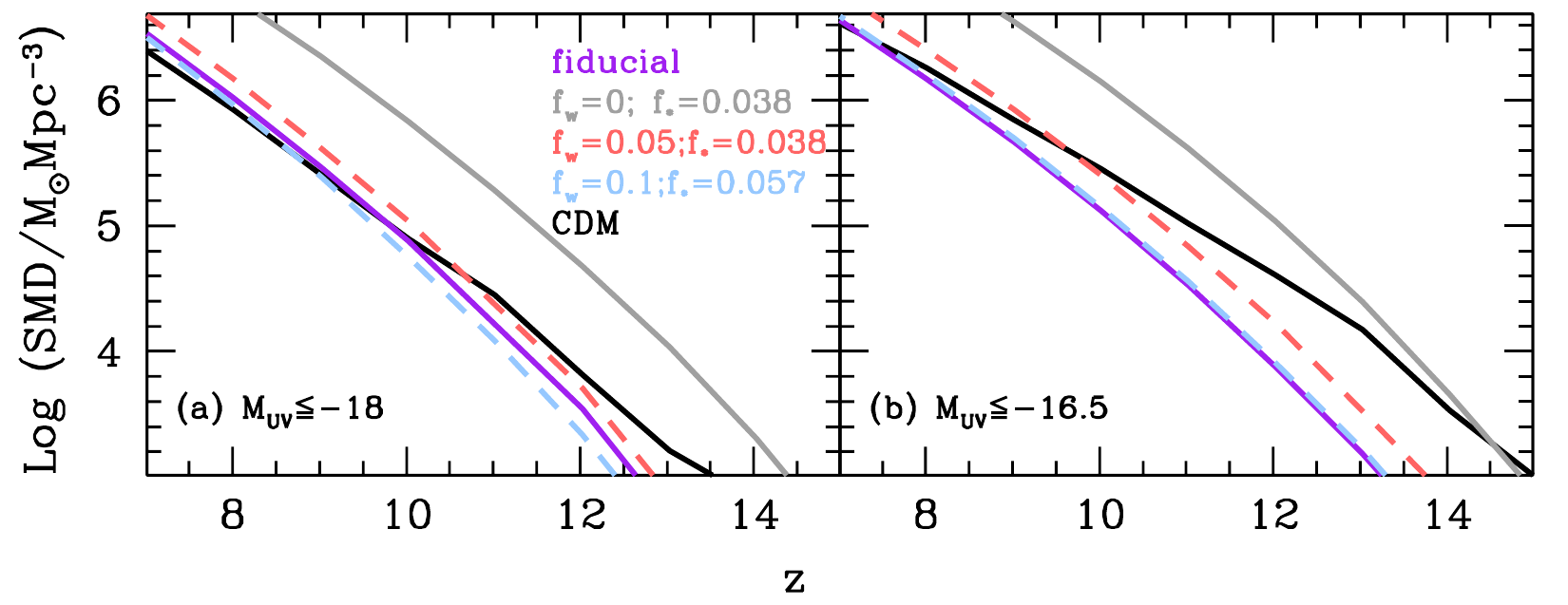} }
  \caption{Redshift evolution of the SMD for the $1.5\kev$ WDM scenario. For reference, in each panel we show results using the fiducial $1.5\kev$ and CDM models (solid purple and black lines, respectively) and a model with no SN feedback in the $1.5 \kev$ WDM scenario (grey line). The red and blue dashed lines show $1.5 \kev$ results obtained by varying the two free parameters (star formation efficiency threshold $f_*$ and the fraction of SN energy driving winds $f_w$) by 50\% as compared to the fiducial model. The left and right panels show the SMD from galaxies that have already been detected, i.e. $\MUV \leq -18$, and galaxies that are expected to be detected using a magnitude limit of $\MUV \leq -16.5$ for the JWST, respectively. As seen, varying the free parameter values only affects the normalization of the SMD with the slope remaining unchanged, specially integrating down to $M_{UV}=-16.5$ (right panel). The slope of the SMD is independent of the astrophysics implemented, presenting a robust observable to distinguish between DM models.  } 
\label{fig_smdpred2} 
\end{figure*}

The SMD, integrated to a fixed UV luminosity threshold, depends both on the number of DM halos and their star formation histories. As shown in the above sections, WDM models with $\mx\lsim 2\kev$ are fundamentally different from colder DM models, having both: {\it(i)} a dearth of low-mass DM halos, and {\it (ii)} a younger, more luminous UV population at a given stellar mass.  Effects (i) and (ii) act in opposite directions.  Younger stellar populations (ii) allow smaller galaxies to be detectable at a given UV threshold, partially compensating for the dearth of corresponding halos (i). Nevertheless, effect (i) dominates, as evidenced by the piling up of $\MUV$ contours near the faint end of the $1.5$ keV WDM SMDs: fainter galaxies require lower mass halos which do not exist.  Hence, the SMD evolution is steeper with $z$ in the $1.5 \kev$ scenario compared to the other three DM models: while galaxies with $\MUV$ as faint as $-10$ contribute to the total SMD in the CDM scenario, the contribution mostly comes from $\MUV \leq -14$ galaxies in the $1.5\kev$ model since smaller galaxies can not form due to the large free-streaming scale of this light DM particle. Indeed, at $z \simeq 9$, 50\% of the SMD comes from galaxies brighter than $\MUV =-17$ in the $1.5 \kev$ WDM model. 

For ease of comparison, in Fig. \ref{fig_smdpred} we show the SMD evolution for all of our DM models, together in the same panels. The SMDs are comparable for all models for galaxies brighter than $\MUV =-18$ up to $z=10$; with this bright limit, the SMD in the $1.5\kev$ case starts showing a steeper evolution at $z >10$, and is 0.4 dex lower than in the three other models at $z\simeq 12.5$. However, the situation changes integrating down by 1.5 magnitudes to $\MUV = -16.5$, which is a reasonable limit for the JWST. With this limit, the $1.5\kev$ SMD evolution becomes steeper than the three other models as early as $z \simeq 8$ and the difference becomes increasingly pronounced with $z$, with CDM  predicting about 0.5 dex (1 dex) more stellar mass per unit volume at $z \simeq 11$ ($z \simeq 13$). The $z$-evolution of the SMD can be parameterised as $\log (SMD) = \gamma (1+z) + \delta$ where we find $(\gamma,\delta) = (-0.44,10.3)$ and $(-0.63,11.9)$ for CDM and the $\mx = 1.5\kev$ WDM, respectively. With its stepper slope, the $z-$evolution of global quantities like the SMD will be instrumental in differentiating the standard CDM from WDM models that invoke particles lighter then $2 \kev$. 

Finally, we show the impact of the free parameter values on the $z$-evolution of the SMD. We re-calculate the SMD in the $1.5\kev$ by varying $f_*$ and $f_w$ by 50\%, in addition to running our model with no SN feedback ($f_w=0$). As shown in Fig. \ref{fig_smdpred2}, varying the two model free parameters only affects the normalization of the SMD. While the slope varies slightly for galaxies brighter than $M_{UV}=-18$ (left panel), {\it the slope maintains its value of $-0.63$} quoted above integrating down to a limit of $M_{UV}=-16.5$. This shows that the slope of the SMD is unaffected by astrophysical uncertainties on integrating down to magnitudes accessible by the {\it JWST}, and offers a robust method of distinguishing between CDM and WDM models.

\section{Conclusions }
The standard $\Lambda$CDM cosmological model has been extremely successful at explaining the large scale structure of the Universe. However, it faces a number of problems on small scales (e.g. the number of satellite galaxies and the DM halo profiles) that can potentially be solved by invoking warm dark matter (WDM) comprised of low mass (\kev) particles. Since WDM smears-out power on small scales, it effects are expected to be felt most strongly on the number densities and the assembly history of the earliest (low mass) galaxies that formed in the Universe. Here we explore how current and upcoming observations of \highz ($z \gsim 7$) LBGs can constrain the mass of WDM particles.

We consider four different DM scenarios: CDM and WDM with $\mx =1.5, 3$ and $5\kev$. Building on DM merger trees, our galaxy formation model includes the key baryonic processes of star formation, feedback from both supernovae (SN) explosions and photo-heating from reionization, and the merger, accretion and ejection driven growth of stellar and gas mass. Below we summarize our main results.

We find that the observed UV LF ($\MUV \lsim -17$) is equally well-fit by all four DM models for a maximum star formation efficiency of $3.8\%$, with 10\% of SN energy driving winds ($f_* = 0.038$ and $f_w = 0.1$). However, the $1.5 \kev$ WDM UV LF starts to peel-away from the other three (which are identical down to $\MUV=-12$ for $z=7-12$) for $\MUV \gsim -16$ at $z \geq 10$. It exhibits a shallower faint-end slope ($\alpha$) by about 0.1-0.3 and and shows a drop of about 0.5 dex in the number density at $\MUV \simeq -15,-16$ at $z =11-12$. Given the small differences, even with its capabilities of constraining the shape of the UV LF down to $M_{UV} \simeq -16$ the JWST will be hard pressed to obtain constraints on whether $\mx \gsim 2 \kev$ or $\mx \lsim 2 \kev$, solely using the UV LF.

The suppression of small scale structure leads to delayed and more rapid stellar assembly in the $1.5\kev$ WDM scenario (compared to the three other models) which results in galaxies of a given stellar mass being more UV luminous, i.e. a lower M/L ratio. While the M/L relation for CDM (and $\mx \geq 3\kev$ WDM) is well-fit by the functional form ${\rm log}\, \M* = -0.38 \MUV + 2.4-0.1z$, the M/L ration for the $1.5 \kev$ WDM starts diverging from this relation at $\MUV =-19$ at $z \simeq 7$, with a $z$-evolution in both the slope and normalisation. The lower M/L ratios in the $1.5\kev$ scenario partially compensates for the dearth of low mass halos, as a result of which the UV LFs predicted by our semi-analytic galaxy evolution model are more similar than simple estimates based on scaling of the halo mass functions.

Finally we estimate the redshift evolution of the SMDs, which provide a more direct probe of the mass assembly history (albeit requiring accurate multi-band photometry).  Integrating down to a limit of $\MUV \simeq -16.5$ (corresponding to a conservative {\it JWST} threshold), we find that the 1.5 keV WDM SMDs evolve more rapidly with redshift than those predicted by CDM. Specifically, we find $\log (SMD) \propto -0.44 (1+z)$ for CDM, with a steeper slope of $\log(SMD) \propto -0.63 (1+z)$ for WDM with $\mx =1.5 \kev$. Indeed, CDM predicts about 3 (10) times more stellar mass per unit volume as compared to the $1.5 \kev$ scenario at $z \simeq$ 11 (13) integrating to magnitudes of $\MUV \simeq -16.5$. We also show that the astrophysical parameters only affect the normalization of the SMD, with the slope being independent of the free parameter values integrating down to magnitudes of $\MUV \simeq -16.5$. \citet{maio-viel2015} have used cosmological hydrodynamical simulations that include sub-grid treatments of star formation and feedback, metal-line cooling and a metallicity dependent initial mass function, in addition to following the detailed chemical enrichment of early galaxies. Interestingly, in spite of their very different approach, these authors find a similar result, namely that global quantities such as the SMD and specific star formation rate will provide powerful probes of the nature of DM at these early cosmic epochs.

To conclude, we find that the build up of observable high-$z$ galaxies is similar in CDM as compared to WDM models with $\mx \geq 3 \kev$. However, structure formation (and hence the baryonic assembly) is delayed and subsequently proceeds notably faster for $\mx \lsim 2 \kev$ than for CDM.  We expect the corresponding rapid redshift evolution of the SMD to be detectable with the upcoming {\it JWST}, providing a powerful testbed for WDM models.

\section*{Acknowledgments} 
PD acknowledges the support of the Addison Wheeler Fellowship awarded by the Institute of Advanced Study at Durham University and of the European Research Council and thanks M. Haehnelt, A. Mazumdar, R. McLure and M. Viel for useful discussions. The authors thank A. Grazian and co-authors for allowing us to use their results and for their instructive comments. 

\bibliographystyle{apj}
\bibliography{ms}

\label{lastpage} 
\end{document}